\documentclass[aps,pra,twocolumn,groupedaddress,showpacs]{revtex4}

\usepackage{fancyhdr}
\usepackage{bm}
\newcommand{\mbf}[1]{\mathbf{#1}}
\newcommand{\mrm}[1]{\mathrm{#1}}
\newcommand{\eref}[1]{(\ref{#1})}
\newcommand{\sigmu}{\sigma}
\newcommand{\wedg}{}

\begin{document}


\title{Constraining Validity of the Minkowski Energy--Momentum Tensor}


\author{Robert N. C. Pfeifer}
\email{pfeifer@physics.uq.edu.au}
\author{Timo A. Nieminen}
\author{Norman R. Heckenberg}
\author{Halina Rubinsztein-Dunlop}
\affiliation{The University of Queensland, Centre for Biophotonics and Laser Science, School of Physical Sciences, Brisbane, QLD 4072, Australia}

\date{\today}

\begin{abstract}
There exist two popular energy--momentum tensors for an electromagnetic wave in a dielectric medium. The Abraham expression is robust to experimental verification but more mathematically demanding, while the Minkowski expression is the foundation of a number of simplifications commonly found within the literature, including the relative refractive index transformation often used in modelling optical tweezers. These simplifications are based on neglecting the Minkowski tensor's material counterpart, a process known to be incompatible with conservation of angular momentum, and in conflict with experimental results, yet they are very successful in a wide range of circumstances. This paper combines existing constraints on their usage with recent theoretical analysis to obtain a list of conditions which much be satisfied to safely use the simplified Minkowski approach. Applying these conditions to an experiment proposed by Padgett \emph{et~al.}, we find their prediction in agreement with that obtained using the total energy--momentum tensor.
\end{abstract}

\pacs{42.25.Bs, 03.50.De, 41.20.Jb\\\ \\Published as: Phys. Rev. A 79(2), 023813 (2009)}

\maketitle

\section{Introduction}

For a hundred years, the energy--momentum tensor of an electromagnetic wave in a dielectric medium has been a subject of debate, and with it, the 
momentum 
and angular momentum 
densities 
of light within the medium. This debate is frequently referred to as the Abraham--Minkowski controversy in honour of two of its earliest contributors.

The first expression proposed for the energy--momentum tensor of an electromagnetic wave in a dielectric was that given by \textcite{minkowski} in 1908, corresponding to a linear momentum density of $\mathbf{D}\times\mathbf{B}$. The Minkowski electromagnetic energy--momentum tensor was criticised for its lack of transpose symmetry, which led \textcite{pauli} to observe that ``Torques [\ldots] appear which cannot be compensated for by a change in the electromagnetic angular momentum''. We now know this asymmetry to be incompatible with conservation of angular momentum 
\cite{jackson}.

An alternative, transpose-symmetric tensor was proposed by \textcite{abraham} in 1909, corresponding to a linear momentum density of $\mathbf{E}\times\mathbf{H}/c^2$. In a recent paper \cite{pfeifer} we reviewed the ensuing debate and concluded that the electromagnetic energy--momentum tensor does not give a complete description of a physical system on its own, and that the total energy--momentum tensor must be considered,
incorporating terms relating to the motion of matter as well as the electromagnetic wave.

This point is well demonstrated by the thought experiment of \textcite{balazs}, who shows that when an electromagnetic wave enters a material medium, the medium must be set in motion in the direction of propagation of the wave if the total linear momentum of the system is to be conserved. We show in \cite{pfeifer} that if the medium is initially at rest then the velocity of this motion is given by
\begin{equation}\label{eq:vdependence}
\mathbf{v}=\frac{1}{\rho_0}\left(n_g - 1\right)\frac{\mbf{E}\times\mbf{H}}{c^2},
\end{equation}
where $\rho_0$ is the matter density in the local rest frame, $\mu$ and $\mu_0$ are the magnetic permeabilities of the medium and of free space respectively, and $n_g$ is the group refractive index of the medium.
Thus to accurately model a physical system incorporating both electromagnetic waves and 
material media,
we must consider
the momentum of the wave and the momentum of the medium, with particular attention to the additional momentum imparted to the medium while it is being traversed by the electromagnetic wave. Similar considerations apply to the discussion of angular momentum.
This position is further substantiated by recent literature such as \cite{loudon3}.
In this paper we present a systematic determination of the circumstances under which each approach may be employed. We concentrate on the energy--momentum tensor formalism --- discussion of Lorentz force approaches may be found in 
\cite{loudon2006}
and references therein.

A recent experimental proposal by \textcite{padgett} aims to study the behaviour of a glass disc as it is traversed by laser pulses carrying orbital angular momentum. The authors calculate momentum transfer to the disc according to the historic Minkowski and Abraham approaches, determining that if the Minkowski approach is applicable here, the disc will remain stationary, and if the Abraham approach is applicable, the disc will rotate. We demonstrate a unified approach in which the total energy--momentum tensor may be partitioned into material and electromagnetic components in accordance with either the Abraham or Minkowski scheme \cite{pfeifer,mikura}, and that both yield the same physical result, which agrees with the historic Abraham approach. We therefore predict that the disc will rotate, in agreement with \textcite{padgett}, who use an Einstein's Box argument, and \textcite{mansuripur2005} and \textcite{loudon2003}, who use a Lorentz force approach (Loudon in fact demonstrates the agreement between these two methods).

What we present is a definitive system for calculating momentum transfer in any physical situation, with guidelines on when the historic Minkowski approach might also be appropriate. Our technique offers concrete predictions for the outcome of the experiment proposed in \cite{padgett}, or indeed for any other experiment involving transfer of electromagnetic linear or angular momentum.
Importantly, the approach is also consistent with previous pivotal experiments employed to attempt to distinguish between the Abraham and Minkowski formulations 
\cite{pfeifer,ashkin,walker,jones}.

Our concern is primarily with the classical regime. However, the Abraham--Minkowski controversy is also relevant to physics on the quantum scale. For example, momentum transfer to charge carriers in a semiconductor \cite{loudon3} is relevant to solar cell technology, and optical momentum transfer is also important in quantum atom optics \cite{campbell}. \textcite{leonhardt2006} provides a thorough treatment of electromagnetic momentum transfer in Bose--Einstein condensates, relating 
the Abraham and Minkowski expressions both to one another and to the behaviour of the condensate. In \cite{leonhardt2006a} he comments that the extension of quantum treatments to the macroscopic world remains unclear. With this 
article
we aim to fill that gap.

\section{\label{sec:totememt}Energy--Momentum Tensors}
\subsection{The total energy--momentum tensor}
The total energy--momentum tensor 
is a four-dimensional second rank tensor which describes the density and flux of energy and momentum within a system. It takes the general form
\begin{equation}\label{eq:totemt}
T=\left(
\begin{array}{cc}
u & \mathbf{S}/c\\
c\mathbf{g} & -\sigmu
\end{array}
\right)
\end{equation}
where \textit{u} is the energy density, \textbf{S} describes the flux of energy and in free space corresponds to the Poynting vector, \textbf{g} is the momentum density, $\sigmu$ describes the flux of momentum and in free space corresponds to the Maxwell stress tensor, and \textit{c} is the speed of light.
Conservation of linear momentum imposes the constraint
\begin{equation}
\partial_\alpha T^{\alpha\beta}=0\label{eq:linmom}
\end{equation}
and conservation of angular momentum then requires \cite{jackson} \footnote{A hypothetical exception is in space--times with torsion, in which conservation of angular momentum would not necessarily impose transpose symmetry.}
\begin{equation}
T^{\alpha\beta}=T^{\beta\alpha}. \label{eq:reflecSym}
\end{equation}
Confirmation of this symmetry was provided by the experiments of James \cite{james,james2} and Walker et~al. \cite{walker,walker2}. Derivations of the canonical energy--momentum tensor frequently lack this symmetry, and it must then be manually reintroduced 
(e.g. \cite{belinfante1940,jackson}).
However, \citeauthor{eddington1924} argues \cite{eddington1924} and more recently \citeauthor{gamboa-saravi2002}
shows \cite{gamboa-saravi2002,gamboa-saravi2004}
that when the canonical energy--momentum tensor 
is formulated in an explicitly covariant manner, 
it will automatically satisfy (\ref{eq:reflecSym}).

We will use a total energy--momentum tensor 
presented by \textcite{mikura}, derived for a nonviscous, compressible, nondispersive, polarisable, magnetisable, isotropic fluid, and including electrostrictive effects, magnetostrictive effects, and acoustic waves. In \cite{pfeifer} we also indicate how the approach may be extended to dispersive media. 
The full expression for the total energy--momentum tensor is given in our earlier paper, but here it suffices to take the nonrelativistic limit, inserting
\begin{eqnarray}
u&=&\begin{array}{l}
\frac{1}{2}\left(\mbf{E}\cdot\mbf{D}+\mbf{H}\cdot\mbf{B}\right)+\rho_0 (c^2+\epsilon_i)\end{array}\\
\sigmu &=& \begin{array}{l}\mbf{E}\wedg\mbf{D}+\mbf{H}\wedg\mbf{B}-\frac{1}{2}\left(\mbf{E}\cdot\mbf{D}+\mbf{H}\cdot\mbf{B}\right)\mbf{I}\end{array}\nonumber\\
&&-\rho_0\mbf{v}\wedg\mbf{v} - \phi\mbf{I}\\
c\mbf{g}=\mbf{S}/c&=&\mbf{E}\times\mbf{H}/c+\rho_0 c \mbf{v}\label{eq:totmomdens}
\end{eqnarray}
into equation (\ref{eq:totemt}),
where $\rho_0$ is the density of the material medium in the local rest frame, $\epsilon_\mrm{i}$ is the specific internal energy of non-electromagnetic nature, \textbf{I} is the identity matrix, \textbf{E}, \textbf{B}, \textbf{D}, and \textbf{H} take their usual meanings in the Maxwell equations, and $\phi$ is the total pressure, which may include electrostrictive and magnetostrictive effects. The 
notation
$\mbf{A}\wedg\mbf{B}$ denotes the dyadic product $(A\wedg B)^{ij}=A^iB^j$.

An expression for the total momentum density associated with an electromagnetic wave in a dielectric medium may be obtained by substituting \eref{eq:vdependence} into \eref{eq:totmomdens}. It may also be obtained by applying conservation of momentum to the propagation of a wave pulse from free space into a dielectric medium, in the absence of acoustic effects (and hence no surface or bulk matter waves), and this is the means by which \eref{eq:vdependence} is derived. The total momentum density associated with the presence of an electromagnetic wave is also called the canonical momentum density \cite{garrison}, and corresponds to the canonical momentum density of Lagrangian dynamics.

\subsection{The Abraham and Minkowski formulations}

The Abraham and Minkowski formulations may be obtained from the total energy--momentum tensor as follows:

\subsubsection{The Abraham formulation}
To obtain the Abraham formulation, we separate the total energy--momentum tensor into two parts, an electromagnetic tensor due solely to the fields themselves, and a material tensor due to the motion of the dielectric medium. 
\begin{eqnarray}
T_{\mrm{EM, Abr}}&=&\left( \begin{array}{cc}
\frac{1}{2}\left(\mbf{E}\cdot\mbf{D}+\mbf{H}\cdot\mbf{B}\right) & \frac{1}{c}\mbf{E}\times\mbf{H} \\
\frac{1}{c}\mbf{E}\times\mbf{H} & -\sigmu_\mathrm{dielectric}
\end{array}\right)\label{eq:AbrEMEMT}\\
T_{\mrm{mat, Abr}}&=&\left( \begin{array}{cc}
\rho_0 (c^2+\epsilon_i) & \rho_0 c \mbf{v}\\
\rho_0 c\mbf{v} & \rho_0\mbf{v}\wedg\mbf{v} + \phi\mbf{I}
\end{array}\right)\nonumber\label{eq:AbrMat}
\end{eqnarray}
where $\sigmu_\mathrm{dielectric}$ is the generalisation of the free space Maxwell stress tensor to dielectric materials,
\begin{equation}
\begin{array}{c}
\sigmu_\mathrm{dielectric}=\mbf{E}\wedg\mbf{D}+\mbf{H}\wedg\mbf{B}-\frac{1}{2}\left(\mbf{E}\cdot\mbf{D}+\mbf{H}\cdot\mbf{B}\right)\mbf{I}.
\end{array}
\end{equation}
In the nonrelativistic limit we assume the material would be at rest in the absence of electromagnetic fields, and consequently $\mbf{v}$ corresponds to the motion induced by the presence of the electromagnetic wave, which depends on the electromagnetic fields in accordance with
(\ref{eq:vdependence}) above. 

It is not a surprise that some of the momentum of a propagating electromagnetic wave is carried by the particles of the medium
 --- the important role of the material counterpart to the Abraham tensor was first recorded by \textcite{jones} in 1954. However, Jones and Richards were unaware of the requirement that the Minkowski tensor also be accompanied by a material energy--momentum tensor in order to conserve angular momentum, and that this tensor is also field-dependent. Their expression for the material counterpart to the Abraham tensor was obtained by requiring that the total momentum density under the Abraham approach
be equal to that 
given by the Minkowski
energy--momentum
tensor 
used without a material counterpart.
Their expression for the Abraham material momentum density is therefore incomplete.

The role of the material medium was subsequently confirmed by \textcite{gordon} who 
explicitly described
material mechanisms of momentum transfer in Jones and Richards' experiment.

\subsubsection{The Minkowski formulation\label{sec:minkform}}

The Minkowski electromagnetic energy--momentum tensor in isolation is incompatible with conservation of angular momentum. The problems caused by its asymmetry have been debated at length \cite{abraham,moller,pauli} and historically formed the primary arguments in favor of the Abraham expression. This is because neither the Abraham nor the Minkowski tensor was initially proposed with a material counterpart.


We have already seen that a material counterpart tensor is necessary for the Abraham tensor to correctly describe transfer of linear momentum. Similarly, we may attribute a material counterpart to the Minkowski energy--momentum tensor to resolve its problems with angular momentum. However, this counterpart was not proposed until experiments were performed by \textcite{james,james2} and \citeauthor{walker} \cite{walker,walker2} in 1968--1977 which appeared to favour the Abraham tensor (with counterpart) over the Minkowski case. A material counterpart
to the Minkowski tensor 
was then proposed by \textcite{israel} to resolve this conflict.

To obtain the Minkowski formulation, we divide the total energy--momentum tensor as follows:

\begin{eqnarray}
T_{\mrm{EM, Mink}}&=&\left( \begin{array}{cc}
\frac{1}{2}\left(\mbf{E}\cdot\mbf{D}+\mbf{H}\cdot\mbf{B}\right) & \frac{1}{c}\mbf{E}\times\mbf{H} \\
c\,\mbf{D}\times\mbf{B} & -\sigmu_\mathrm{dielectric}
\end{array}\right)\label{eq:MinkEMEMT}\\
T_{\mrm{mat, Mink}}&=&\left( \begin{array}{cc}
\rho_0 (c^2+\epsilon_i) & \rho_0 c \mbf{v}\\
\left[\begin{array}{c}
\rho_0 c\mbf{v}-c\,\mbf{D}\times\mbf{B}\\+\frac{1}{c}\mbf{E}\times\mbf{H}
\end{array}\right]& \rho_0\mbf{v}\wedg\mbf{v} + \phi\mbf{I}
\end{array}\right).\nonumber\label{eq:MinkMat}
\end{eqnarray}

Note that $\mbf{v}$ is defined identically for both the Abraham and Minkowski formulations, and  the additional terms in the Minkowski ``material'' energy--momentum tensor do not correspond to a physical motion of the material medium. This is required by the experiments of \citeauthor{james} and \citeauthor{walker} 

\subsection{The simplified Minkowski approach}

\subsubsection{Comparison with the Abraham approach\label{sec:compar}}

The rationale for the Minkowski formulation may at first appear obscure. Nevertheless, it is the basis of several convenient techniques which can be used to simplify calculations. Frequently these techniques involve neglecting the material counterpart tensor, which is puzzling as the momentum flux obtained from this tensor seldom goes to zero. Further consideration will reveal why these techniques often work, and the range of circumstances under which they can be employed.

In Sec.~VIII.B of \cite{pfeifer} we described a number of different types of experiment in which momentum transfer may take place. In particular, we considered momentum flux across a boundary (Sec.~VIII.B.2), and the coupling of momentum into a physical system when that system is immersed in a penetrating electromagnetic wave (Sec.~VIII.B.1). However, we did not address a further important distinction, which is whether there exists an ongoing momentum flux between the electromagnetic wave and the apparatus in the steady state condition (for example where momentum transfer from the beam to a mirror is in equilibrium with an external force), or whether momentum transfer only occurs as the beam enters or leaves the medium. In experiments involving a momentum flux at steady state, simplified approaches based on neglecting the Minkowski  material tensor are effective. In those where momentum flux at steady state vanishes, they fail, and systems of this category are used to confirm the form of the total energy--momentum tensor, for example the experiments of James, and Walker et~al.

Systems which exhibit momentum flux at equilibrium include reflection experiments (e.g. \cite{jones,jones2}), and refraction experiments, such as optical tweezers \cite{ashkin1986,nieminen2004d}. Reflection is especially simple, as the rate of change of momentum of the mirror depends entirely on the momentum flux 
across its boundary. To illustrate how the simplified Minkowski approach works in such systems, let us consider as an example a laser beam reflecting perpendicularly off a mirror which is suspended in a dielectric.

There are two beams present in this experiment --- the incoming and the reflected beam. They exist in superposition, and the velocities which they induce in the dielectric medium are equal and opposite. The medium therefore remains stationary, and we can ignore terms in $\mbf{v}$. 
The rate of change of momentum density within the mirror is given by
\begin{equation}
\partial_t\mbf{g}^i_\mrm{mirror,Abr} = -\frac{1}{c}\partial_t T^{i0} + \frac{1}{c}\partial_j T^{ij}\quad\ 1\leq(i,j)\leq3,\label{eq:boundaryflux}
\end{equation}
where $T$ at the surface of the mirror is discontinuous, and must be treated as the limit of a continuous expression.
We then apply Gauss's Law to obtain Eq.~(53) of \cite{pfeifer}, which tells us how to calculate the force per unit area on the mirror:
\begin{equation}
\mbf{F}^i_B=-\oint\mrm{d}S_j\, \sigmu^{ij}_{A,\mrm{in}}-\oint\mrm{d}S_j\, \sigmu^{ij}_{B,\mrm{out}}.\label{eq:boundary}
\end{equation}
Here, $\sigmu^{ij}_{A\,\mrm{in}}$ is the stress tensor for the inbound wave just outside the mirror. Technically, $\sigmu^{ij}_{B\,\mrm{out}}$ represents the component of the stress tensor just inside the mirror responsible for transferring momentum to the outbound wave, but by conservation of momentum it suffices instead to consider the negative of the stress tensor for the outbound wave just outside the mirror. 
Performing a plane wave decomposition at the surface of the mirror
\begin{eqnarray}
\mbf{E}(x,t,\omega)&=&{E_0}(\omega)\mrm{e}^{\mrm{i}(\omega t - n_\phi(\omega) k x)}\mbf{\hat e_1}\\
\mbf{H}(x,t,\omega)&=&{H_0}(\omega)\mrm{e}^{\mrm{i}(\omega t - n_\phi(\omega) k x)}\mbf{\hat e_2}\\
\frac{{E}_0}{{H}_0} = &Z& = \sqrt{\frac{\mu(\omega)}{\epsilon(\omega)}}
\end{eqnarray}
where $\mbf{\hat e_1}$, $\mbf{\hat e_2}$, and $\mbf{\hat e_3}$ are orthogonal unit vectors with $\mbf{\hat e_3}$ normal to the surface of the mirror,
and substituting into $\sigmu^{ij}$, we readily find
\begin{equation}
\frac{\mbf{F}_\mrm{Abr}}A=2n_\phi{\mbf{E}\times\mbf{H}}/{c}.\label{eq:fperaA}
\end{equation}
This is the expression obtained using either the Abraham tensor with material counterpart or the Minkowski tensor with counterpart, as both add up to the same total energy--momentum tensor.

We now consider the Minkowski electromagnetic energy--momentum tensor in isolation. The momentum density
\begin{equation}
\mbf{g}_\mrm{Mink}=\frac{1}{c}T^{i0}=\mbf{D}\times\mbf{B}
\end{equation}
is increased by a factor of $n_\phi^2$ relative to the Abraham expression $\mbf{E}\times\mbf{H}/c^2$. The Maxwell stress tensor we presented in \eref{eq:MinkEMEMT}, however, remains unchanged, and if we use this to evaluate the force per unit area we will of course obtain the same expression as above. In practice, however, the usual procedure is to infer the rate of momentum transfer from the speed of wave propagation and 
the momentum density. In other words, we should apply conservation of linear momentum \eref{eq:linmom} neglecting the material medium, to redefine $\sigmu^{ij}$ according to
\begin{eqnarray}
\partial_j \sigmu^{ij} = c\,\partial_t \mbf{g} &=& \partial_t (c\,\mbf{D}\times\mbf{B}).
\end{eqnarray}
This would give a very different result from \eref{eq:fperaA}, except that
instead of using \eref{eq:boundary}, we will now (incorrectly) assume that the momentum flux at steady state is given by the momentum density multiplied by the rate at which a wavefront would propagate through the medium:
\begin{eqnarray}
\frac{\mbf{F}_\mrm{Mink}}{A}&\stackrel{!}{=}&2\frac{c}{n_g}\mbf{D}\times\mbf{B}\label{eq:fperaM1}\\
&\stackrel{!}{=}&2\frac{n_\phi^2}{n_g}\mbf{E}\times\mbf{H}/c. \label{eq:fperaM}
\end{eqnarray}
We have placed an exclamation mark over the equality as a reminder to the reader that this expression is not physically valid.

This is 
a case of two wrongs very nearly making a right, and occurs
whenever a simplified Minkowski approach is successfully employed. As well as in the explicit implementation described above, this also takes place when the simplified Minkowski approach is employed implicitly, for example in the relative refractive index technique 
often employed in modelling optical tweezers. In this technique, a particle of refractive index $n_\mathrm{particle}$ in a liquid dielectric medium of refractive index $n_\mathrm{medium}$ is treated as a particle of refractive index $n_\mathrm{particle}/n_\mathrm{medium}$ in vacuum.

No such lucky cancelling of errors occurs for experiments of the sort described by \textcite{james,james2}, or Walker et~al. \cite{walker,walker2}. In these experiments an electromagnetic wave traverses a medium essentially unchanged. Conservation of momentum arguments reveal that the medium must move with a steady average velocity while being traversed by the electromagnetic wave \cite{balazs}, so there is only a net momentum flux while the medium is being traversed by the leading or trailing edge of the 
wave, and these two impulses are equal and opposite. For the intervening time in which the medium is entirely immersed in a steady state electromagnetic wave, or vice versa, no momentum transfer takes place (neglecting absorptive behaviours such as opto-acoustic coupling).

Consequently, in this sort of experiment, 
we are dealing in the steady state not with momentum flux but with momentum densities. The behaviour of the material medium is calculated from the $\mbf{v}$-dependent terms of the material component of the material momentum density. In contrast, the simplified Minkowski approach would assume that all momentum not contained in $\mbf{D}\times\mbf{B}$ is material. Because there is no momentum flux in the steady state calculation, there is no opportunity in this situation for a second error of the sort described above \eref{eq:fperaM1} to cancel out the error in momentum density, which comes from treating the entire Minkowski ``material'' momentum density as describing the motion of the material medium.

These experiments 
directly probe the electromagnetic momentum density, and serve to confirm that the wave portion of the total energy--momentum tensor does take the form \eref{eq:AbrEMEMT}. This confirmation motivated the introduction of a specific material counterpart tensor to the Minkowski electromagnetic energy--momentum tensor \cite{israel}.

We conclude that the simplified Minkowski approach cannot be used when the behaviour of interest results solely from transit of an electromagnetic wave through a medium without absorption, reflection, or refraction. Because it requires that second error, which arises in calculation of the momentum flux, it is only useful in calculating momentum fluxes and not momentum densities.

The field terms of the total energy--momentum tensor are confirmed by the experiments of James, and Walker et~al., but the cautious reader might also ask how confident we are in our derived expression for momentum flux \eref{eq:fperaA}, which differs from \eref{eq:fperaM} only by a factor of $n_\phi/n_g$. This result has in fact also been experimentally verified, through the reanalysis by \textcite{garrison} of Jones and Leslie's experiment of \citeyear{jones2} \cite{jones2}. In \citeauthor{garrison}'s terminology, expression \eref{eq:fperaA}, which is derived from the total energy--momentum tensor, corresponds to the canonical momentum, while \eref{eq:fperaM} corresponds to the Minkowski momentum. Finally, what \citeauthor{garrison} call the Abraham momentum corresponds to momentum transfer using the Abraham expression for momentum density, but the erroneous momentum flux calculation employed in obtaining \eref{eq:fperaM}. Their finding that momentum transfer is governed by the canonical momentum therefore corresponds to confirmation of Eq.~\eref{eq:fperaA}.

\subsubsection{Limitations on validity}

Based upon what we now know of the simplified Minkowski approach, we may infer the following
conditions under which it may safely be used:
\begin{enumerate}
\item The dynamics of the medium must not be of interest: The simplified Minkowski approach neglects the material energy--momentum tensor, and so approaches based upon it do not reliably model motion of the material medium, or effects such as electro- or magnetostriction. Similarly, as flows within a moving medium may significantly contribute to momentum transfer, the simplified Minkowski approach should only be used when the dielectric medium has time to reach equilibrium \cite{gordon} or when it can be shown that such flows are unimportant.
\item The calculation being performed must be intended to calculate momentum flux, not momentum density, as discussed in Sec.~\ref{sec:compar}.
\item Dispersion must be negligible, such that the factor $n_\phi/n_g$ in \eref{eq:fperaM} goes to 1 (see also \cite{nelson,stallinga}). 
\end{enumerate}
Constraint 3 is not absolute, and in light of Eqs.~\eref{eq:fperaA} and \eref{eq:fperaM}, may be circumvented by multiplying all forces calculated using the simplified Minkowski method by a factor of $n_g/n_\phi$. 

Although the Minkowski electromagnetic energy--momentum tensor has frequently been criticised for incompatibility with conservation of angular momentum, the simplified Minkowski approach may safely be used in experiments involving the flux of angular momentum, provided the above three conditions are adhered to. Once again, the Minkowski expression overestimates the angular momentum density by a factor of $n_\phi^2$ (giving, following \textcite{padgett}, an angular momentum per photon of $n_\phi \hbar k$ as opposed to $\hbar k/n_\phi$), but the simplified flux calculation introduces a compensating error of $(n_\phi n_g)^{-1}$ and the correct result is obtained, up to a factor of $n_\phi/n_g$.

Note that the simplified Minkowski approach is also always valid in the limit $(n^2_\phi-1)\rightarrow 0$, in which the field component of the Minkowski material energy--momentum tensor goes to zero, but this solution is uninteresting as it corresponds to wave propagation in vacuum, in which the Minkowski and Abraham formulations 
trivially 
coincide.
The Minkowski electromagnetic energy--momentum tensor may also be useful in its own right for describing
conservation of pseudomomentum in dielectric media, and this subject is discussed at length elsewhere in the literature \cite{pfeifer,gordon,peierls3,nelson,stallinga}.

\section{Examples\label{sec:padgett}}



As mentioned above, the simplified Minkowski approach is well suited to describing experiments in which an ongoing momentum flux takes place at steady state.
Consequently it should come as little surprise that it predicts the results of \textcite{jones} and \textcite{jones2} with a high level of accuracy --- even more so when the correction for dispersion is applied.
For a suitably reflective mirror the medium is at rest in the steady state (see Sec.~\ref{sec:compar} above); the force on the mirror is dependent upon the momentum flux, and with the correction for dispersion, the result is expected to be in agreement with that obtained from the total energy--momentum tensor.


For optical tweezers, a similar situation holds, with a laser beam being refracted through a dielectric object. This time the inbound and outbound beams are not parallel, but if we neglect absorption, then by symmetry the velocity of the medium adjacent to the particle can only be parallel to its surface. Again we can neglect the medium and satisfy conditions 1--3 above, and the simplified Minkowski approach is a good model of the restoring force when the particle is displaced from the beam's focus.

The simplified Minkowski approach fails when applied to the experiments of Walker et~al. \cite{walker,walker2}, and this is unsurprising as these experiments in effect measure the material momentum density with the electromagnetic wave intensity at steady state. The simplified Minkowski approach is not suited to these experiments, relying as it does on simultaneous errors in the expressions for momentum density and momentum flux.

The experiment of \textcite{ashkin} is an interesting exception. Here, a laser beam passes through a glass box filled with water, and momentum transfer to the water causes the surface to form either a positive or negative lens. As this experiment deals with a laser beam traversing a dielectric medium without reflection or absorption, and with only radially symmetric refraction, we might expect the simplified Minkowski approach to perform poorly. Instead, it performs surprisingly well, and we must turn to a paper by \textcite{gordon} to understand why. By considering explicitly the physical behaviour of the medium, Gordon shows that in these specific circumstances, the dynamics of the boundary will be independent of the formulation adopted. However, the total energy--momentum tensor tells us that the true field momentum density within the medium is given by the Abraham expression, and that of the two fluid dynamical behaviours described by Gordon, the one corresponding to the Abraham approach best reflects the real experiment. Note that in his analysis, Gordon implicitly employed the erroneous calculation of momentum flux in the Minkowski picture which we described in Sec.~\ref{sec:compar}. One cannot help but wonder if once again, this error is responsible for the simplified Minkowski calculation providing the correct result.

Although a similar result should hold for other physically comparable systems, such as a laser beam passing through a solid block \cite{peierls}, one should be wary of using the simplified Minkowski approach in such systems until the means by which it obtains the correct result in them is better understood. In addition, the correction $n_\phi/n_g$ allowing extension to dispersive media has not been demonstrated for such systems and should be avoided.

These historic experiments were key milestones in the development of our current understanding of the Abraham--Minkowski controversy, and the interested reader may find more detail, historical context, and discussion of their contribution to our current understanding of the problem in our recent review paper \cite{pfeifer}. 

Finally, we turn our attention to a new experiment, proposed by \textcite{padgett}, which
consists of a circular glass disc through which 
laser pulses
carrying orbital angular momentum are passed. While a pulse is traversing the disc, some portion of its orbital angular momentum may possibly be transferred to the disc.

These authors demonstrate that if linear electromagnetic momentum transfer is assumed to depend upon $\mathbf{E}\times\mathbf{H}/c^2$, the Abraham expression, then the orbital electromagnetic angular momentum transferred to the disc per photon will go as $\hbar k/n^2$, where $\hbar k$ is the orbital angular momentum per photon in free space. This corresponds to an electromagnetic angular momentum density within the disc of $\bm{\ell}_0/n$, where $\bm{\ell}_0$ is the angular momentum density of the laser pulse in free space, and $n$ is the refractive index of the disc, where no distinction is made between $n_\phi$ and $n_g$.
By conservation of angular momentum, the glass disc therefore acquires an angular momentum of $(1-1/n)\bm{\ell}_0V$ while being traversed by a laser pulse, where $V$ is the volume of the pulse lying within the disc, and this causes it to rotate. 
Electromagnetic torque is only generated while $V$ is changing, so the calculation is similar whether a pulse or a steady state beam is employed, but viscous damping will rapidly bring the disc to rest for a steady state beam.

On the other hand, if the linear electromagnetic momentum density goes as $\mathbf{D}\times\mathbf{B}$, which is the Minkowski expression, then the electromagnetic angular momentum per photon is independent of the refractive index of the medium. The angular momentum of the material medium through which the beam passes, i.e. the disc, therefore remains zero. This corresponds to an increase in the angular momentum density of the beam from $\bm{\ell}_0$ in free space to $n\bm{\ell}_0$ within the disc due to the lower speed of light within the disc.

The behaviour of this experiment is dependent on momentum density while the beam pulse traverses the disc, and not on momentum flux. By our arguments above, the simplified Minkowski approach is unsuitable for use here, a conclusion which is also borne out by consideration of the total energy--momentum tensor:

The angular momentum density predicted for the Minkowski material counterpart tensor is zero, but this does not necessarily correspond to the actual motion of the material medium, which is given by taking only the term in (\ref{eq:MinkMat}) which involves $\mbf{v}$. This value is identical to that obtained under the Abraham expression, and consequently we expect the disc to rotate. If, as here, the motion of the medium is of interest it is vital to recall that the physical momentum of the material medium and the momentum described by the Minkowski ``material'' energy--momentum tensor are not the same, and for this a total (or Abraham) energy--momentum tensor-based approach must be employed.

Can a treatment like that of Gordon's \cite{gordon} for the experiment of \textcite{ashkin} be used here? The answer would appear to be no. 
For linear momentum, regardless of what approach is employed, the momentum of the material medium is never unchanged. It may be of the wrong sign and magnitude, but an interaction between the material medium and the electromagnetic wave exists, so the result may be rectified by some suitable corrective procedure. However, in the simplified Minkowski approach, the angular momentum remains entirely within the electromagnetic wave. There is no transfer to the material medium whatsoever, and hence no finite correction can bring about the correct result. The medium is (erroneously) perceived to be entirely insensible to the angular momentum of the traversing beam.

Finally, we note that the torques experienced by the disc as the light pulse enters and exits are of equal magnitude and opposite direction (if we neglect absorption), and hence the disc will only rotate during the short interval while it is being traversed by the pulse. This will, however, result in observable rotation provided the time required for the pulse to traverse the disc is less than the time for viscous damping forces to bring the disc to rest.

\section{Discussion\label{sec:conc}}

The Abraham--Minkowski controversy has implications for the transfer of both linear and angular momentum between electromagnetic waves and material media. While the Abraham approach is now widely considered to be the most rigorous, simplifications based upon the Minkowski approach remain popular, and although specific experiments are known for which the simplified Minkowski approach breaks down \cite{james,james2,walker,walker2}, a comprehensive treatment of when the approach may safely be employed has been lacking.

We provide three criteria, the satisfaction of which is both necessary and sufficient for the simplified Minkowski approach to correctly model a physical situation. In addition, our comparison of the total energy--momentum tensor approach and simplified Minkowski approach (lacking a material energy--momentum tensor) gives rise to a correction enabling the simplified Minkowski approach to be extended to 
dispersive media
.

Finally, we demonstrate compatibility of our approach with the results of a number of key historic experiments, and apply our guidelines to a new experiment proposed by \textcite{padgett}. This experiment involves the possible transfer of orbital angular momentum from a laser beam to a glass disc, and we predict that rotation of the disc will be observed, with the total energy--momentum tensor formalism being in agreement with the historic Abraham approach.

\acknowledgements

The work of R.N.C.P. was partly supported by 
an Endeavour International Postgraduate Research Scholarship funded by 
the Australian Government Department of Education, Science and Technology.
The authors would like to thank the referees, whose searching questions helped to greatly improve this paper.


\end{document}